# Friend Recommendation based on Hashtags Analysis

Ali Choumane[1], Zein Al Abidin Ibrahim[1]

[1] LARIFA, Faculty of Sciences, Lebanese University, Nabatieh, Lebanon
Email : ali.choumane@ul.edu.lb, zein.ibrahim@ul.edu.lb

**ABSTRACT**

Social networks include millions of users constantly looking for new relationships for personal or professional purposes. Social network sites recommend friends based on relationship features and content information. A significant part of information shared every day is spread in Hashtags. None of the existing content-based recommender systems uses the semantic of hashtags while suggesting new friends. Currently, hashtags are considered as strings without looking at their meanings. Social network sites group together people sharing exactly the same hashtags and never semantically close ones. We think that hashtags encapsulate some people interests. In this paper, we propose a framework showing how a recommender system can benefit from hashtags to enrich users' profiles. This framework consists of three main components: (1) constructing user's profile based on shared hashtags, (2) matching method that computes semantic similarity between profiles, (3) grouping semantically close users using clustering technics. The proposed framework has been tested on a Twitter dataset from the Stanford Large Network Dataset Collection consisting of 81306 profiles.

## 1. Introduction

Recently, social networking websites such as Facebook, Flickr, MySpace, Twitter, etc. have been noticed a rapid growth in the number of registering members. For example, Twitter counts more than 500 million users and about 350K tweets sent per minute[1]. The users need to improve their connections in the social networks, by having new links with others or by being new members in groups or pages of interests. Currently available social networks automatically recommend people to help users find known contacts and discover new relationships. The recommendation is either based on the network relationships (graph topology of the network) or on the content information (interests, skills, shared posts, etc.). The first kind of approach is better at finding known contacts whereas the second ones are stronger at discovering new friends [1]. Relationship-based approach estimates some features in the graph such as the number of common friends. It suggests friends having the highest numbers of mutual friends. Other features are used like the distance between users in the network graph. Content information-based approach computes similarities between users while taking into account their shared information and profiles' attributes and suggests the top k similar users.

Regardless of the method used for recommendation, users need accurate recommendation to help them developing their own personal networks or businesses as well as social network sites are used also for marketing purposes. Nowadays, it is rare to publish posts on social sites without citing *Hashtags* in order to highlight an idea, topic or event. None of the existing recommender systems integrate the semantic of hashtags in the similarity computation between users which constitutes the main contribution of this paper.

The paper is structured as follows: Section 2 shows existing works in the domains of the social information retrieval and recommender systems. Section 3 shows the architecture of the proposed framework consisting of three main components while sections 4, 5 and 6 explain in depth each of these components. Hence, section 4 discusses the need for an accurate hashtag segmentation method and shows our proposition in this way. Section 5 shows how we compute the similarity between profiles and section 6 shows some experiments using clustering to identify k-nearest profiles. Finally, section 7 concludes the paper and proposes some future works.

## 2. Related Work

Social network sites have introduced new communication way by allowing people from diverse areas to meet, interact, share interests and ideas, etc. This encourages a huge number of users to join and reap the potential benefits provided by them [2]. The user generated content poses a challenge in term of information retrieval but presents an advantage for recommender systems. Far away from social networks, recommender systems emerged as an independent research area in the mid-1990's. The recommendation problem is mainly reduced to the problem of estimating ratings for the items that have not been seen by a user. This estimation is usually based on the ratings given by this user to other items and on some other information. Once we can estimate ratings for the yet unrated items, we can recommend to the user the items with the highest estimated ratings [3].

---
[1] http://www.internetlivestats.com/twitter-statistics/

Recommender systems can be classified into three categories, based on how recommendations are made [4]:
1. Content-based recommendations: the user is recommended items similar to the ones the user preferred in the past;
2. Collaborative recommendations: the user is recommended items that people with similar tastes and preferences liked in the past;
3. Hybrid approaches: these methods combine collaborative and content-based methods

Recently, there has been increasing interest toward developing recommender systems for social network sites namely social recommender system, with the aim to suggest information such as blogs, news, web pages, images, tags or individuals [1] by exploiting social network information that are likely to interest users.

Individual recommendation, also known as friend recommendation, represents the main concern of this paper. Several researchers work on this topic. [5] tried to identify missing links in the social network graph. They made recommendation of friends by considering the graph topology, such as computing common neighbors between users. [2] proposed a collaborative filtering framework to facilitate users in exploring new friends based on their interaction intensity and attribute similarity. [1] evaluated two categories of algorithms for recommending people: the first category is based on social relationship information from the social network graph (number of common friends, for example) while the second category is based on content similarity taking into account common keywords between users. They showed that relationship based algorithms outperform content similarity ones in terms of user response. We think that the existing content-based recommender systems have not fully exploited the user's profile information while computing similarities. As mentioned in the previous section, hashtag becomes one of the most popular communication practice in social networks and are used to highlight ideas, topics or events. Therefore, in order to overcome the misperformance of content-based recommender systems, we suggest to integrate hashtag meaning in the similarity computation. In this paper, we propose a system that computes the similarity between profiles by using only the hashtags. This can be considered as an important attribute and can be integrated in any content-based recommender system.

## 3. Proposed Framework

The main idea of this paper concerns the building of a recommender system that helps users finding new relationships on social networks according to their profiles. This system is not based on existing relationship properties of a social network graph, such as the number of common friends or the raw distance between users in the social graph. Our system is based on the information content and more precisely the semantic of cited hashtags.

In social networks, each user is represented by her/his profile which may contain personal information such as the user's name, email, address, age, hobbies, skills, posted texts, images and videos, friend list, etc. The FOAF project defines a set of relevant subjects and properties related to user's profiles. Researchers consider these properties to develop social information retrieval and recommender systems. As mentioned in the previous section, the semantic content of the user's textual posts is not fully taken into consideration. In fact, users cite *hashtags* in their posts to highlight a special meaning about an event or a topic of interest. A hashtag is a word or an un-spaced phrase prefixed with the hash character, #, to form a label. This phrase can be a single word, an acronym, or multiple words joined, and usually identifies the topic of the user's post.

A hashtag allows social sites to group similarly tagged messages, and the retrieval of messages containing hashtags. For example, Instagram, Facebook and Twitter allow users to input a specific hashtag to search all the posts containing it, by exact syntactic matching without looking into the meaning or the words composing the hashtag. In fact, we think that users citing the hashtag #googleabout are somehow interested in Google Company or Google products. The hashtag #androidgames shows that one of the user's interests is android games. A recommender system that explores hashtag meanings would be able to suggest new android games to users citing the hashtag #androidgames. Such a system would also suggest new relationships of people sharing interest to android games although they have not cited the same hashtag. Some other single word hashtags like #shopping, #christmas, etc. give also an idea about the topics of interest of the users citing them.

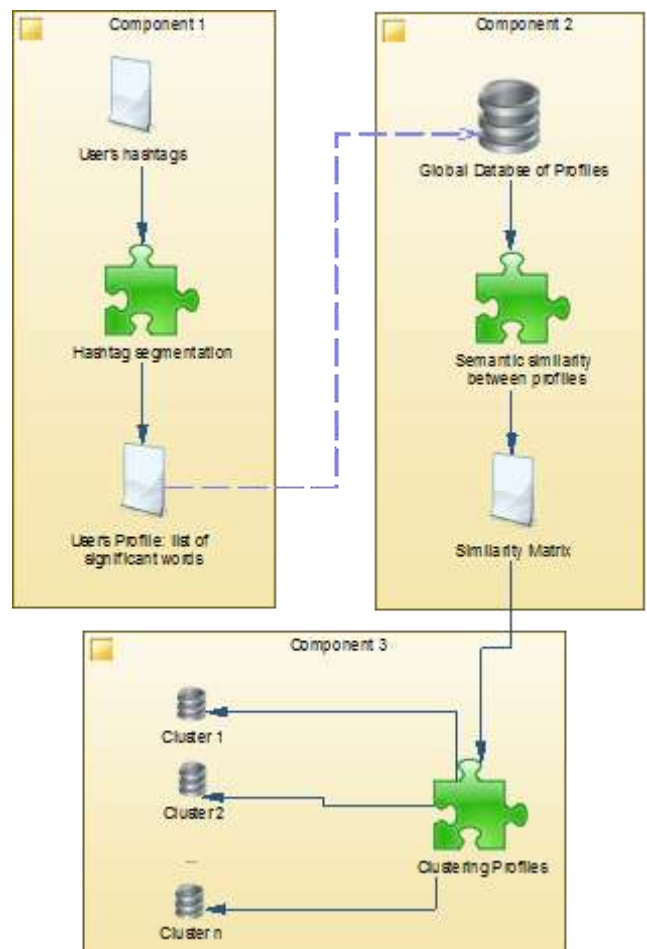

Figure 1: Framework Architecture

Figure 1 shows the different components of our framework. In the component *"Component 1"* of this framework, we consider building for each user a profile based only on the hashtags she/he cited. This profile is complementary to the FOAF profile. By constructing a hashtag-based profile, we mean that the different significant tokens that compose a hashtag should be extracted and added to the user's profile (cf. section 4). The component *"Component 2"* concerns the semantic similarity measures between profiles that allows producing a similarity matrix. Each element of the similarity matrix contains a measure of similarity between two profiles (cf. section 5). In *"Component 3"* we apply a clustering algorithm in order to produce a set of clusters each containing a set of semantically related profiles (cf. section 6). These clusters are the basis of our recommender system, i.e. this gives the possibility to recommend to each user some potential relationships from the cluster she/he belongs to.

## 4. Hashtag Segmentation

A common practice in current social networks is to identify the subjects of a post by means of hashtags, e.g., #Mancherster, #LiesPeopleAlwaysTell, #toobad, #ff, #skypeisnotworkingagain [6].

As defined above, a hashtag is a word or an un-spaced phrase prefixed with the hash character. Hence, a hashtag can be made up of one, two, or more words. In order to use a hashtag, it should be decomposed into its composing words. As much as the number of words increases as much as the complexity of this hashtag and the difficulty of segmenting it into the exact composing words increase. For instance, suppose that we have the hashtag *#dependentrelationship*, this hashtag can be split as *dependent relations hip,* as *dependent relationship,* or as *dependent relation ship.* How to decide what is the right or the most likely segmentation?

Same problem arises with the hashtag *#airportend* that can be split as *air portend*, or as *airport end*, and also as *air port end.*

In our work, we developed a segmentation algorithm that proceeds on two main steps:
1. The first step uses an English lexicon to find all the possible sequences of words that may compose a hashtag. For example, the hashtag *#throwbackthursday* has two lexically correct sequences:
   - *throwback thursday*
   - *throw back thursday*

   This is a lexical step that allows eliminating any segmentation with invalid words, i.e. not found in the dictionary. To accomplish this step, we used the English Lexicon Project[2] made by Washington University consisting of 80000 words [7]. Note that sometimes the hashtag itself is a valid word in the dictionary and added as possible segmentation, for example the hashtag *#worldwide* has two possible segmentations, according to the dictionary: *world wide* and *worldwide*. In this case, we choose the single word as the right segmentation.

2. If at least two possible segmentations arise from the first step, we proceed with a disambiguation step in order to find the most probable sequence of words. We developed a probabilistic model based on bigram frequencies. Note that an n-gram is a contiguous sequence of n items from a given sequence of text or speech [8]. The items can be phonemes, syllables, letters, or words according to the application. In our context, we consider word items. An n-gram of size 2 (n=2) is a bigram. Several corpuses exist and provide bigram frequency counts. We used the bigram list provided by the Corpus of Contemporary American English (COCA)[3]. For each bigram in this list, we computed its probability representing how much this bigram is likely to appear in an English sentence.

To find the most probable segmentation of a hashtag, we consider that each generated segmentation is represented by a path in a Markov model. We select the segmentation with the highest path probability, i.e. the highest product of probabilities along the path.

Consider the hashtag *#worldwidefestival* in order to illustrate this step.

The lexical segmentation step produces the following possibilities:
- *worldwide festival*
- *world wide festival*

The segmentation *worldwide festival* has the bigram probability of 0.0022. The probability of segmentation *world wide festival* is equal to probability of the bigram *world wide* multiplied the probability of the bigram *wide festival* which is 0.05 x 0.0099 = 0.00049. Hence, the segmentation *worldwide festival* is produced.

To evaluate the hashtag segmentation algorithm, we selected the top 387 hashtags trending on social networks in January 2015. We performed an offline segmentation leading to 97.9% success rate. This means that only 8 hashtags are not correctly segmented. Looking in details, we noticed that the corresponding bigrams of 3 hashtags are not found in the COCA corpus. In the other 5 cases, the lexical step failed because the hashtag words are not found in the used English dictionary.

## 5. Profiles Matching

Given the set of cited hashtags of a user, we are now able to derive her/his profile consisting of the different significant words composing these hashtags (cf. section 4). In this section, we show our profiles matching algorithm used to determine whether or not any two profiles share common topics of interest. The algorithm we propose is a generic matching algorithm that measures the semantic similarity between any two profiles. It is generic because it is designed to measure the similarity between any two set of words, not necessarily user's profiles. Such an algorithm could be used to extend our framework to images and videos

---

[2] English Lexicon Project : elexicon.wustl.edu

[3] http://www.ngrams.info/

recommendations as well as such items are usually described by keywords and hashtags.

In order to compute the semantic similarity between profiles, we introduce first the semantic similarity between words.

### 5.1 Semantic similarity between words

Semantic similarity relates to computing the similarity between conceptually similar but not necessarily lexically similar terms. In our framework, we used a similarity measure based on WordNet [9]. WordNet is a large lexical database developed at Princeton University. It attempts to model the lexical knowledge of a native speaker of English. WordNet can also be seen as ontology for natural language terms. It contains around 100,000 terms, organized into taxonomic hierarchies. Nouns, verbs, adjectives and adverbs are grouped into synonym sets (synsets). The synsets are also organized into senses (i.e., corresponding to different meanings of the same term or concept). The synsets (or concepts) are related to other synsets higher or lower in the hierarchy by different types of relationships. The most common relationships are the Hyponym/Hypernym (i.e., Is-A relationships), and the Meronym/Holonym (i.e., Part-Of relationships). For example, taxonomic hierarchies in WordNet allow deriving that the term "feather" is semantically related to the term "bird" as well as the term "bus" to the term "train", etc. Several methods for determining semantic similarity between terms have been proposed in the literature and most of them have been tested on WordNet [10]. Each of the existing measures takes two WordNet concepts c1 and c2 (i.e., word senses or synsets) as input and returns a numeric score that quantifies their degree of relatedness.

The existing semantic similarity methods are classified into four main categories [11]:
1. *Edge Counting Methods*: Determine the similarity between two concepts as a function of the length of the path linking the terms and on the position of the terms in the taxonomy.
2. *Information Content Methods*: Measure the difference in information content of the two concepts as a function of their probability of occurrence in a corpus. More general concepts with many hyponyms have less information content than more specific terms with less hyponyms.
3. *Feature based Methods*: Measure the similarity between two terms as a function of their probabilities or based on their relationships to other similar terms in the taxonomy. Common features lead to increase the similarity and vice versa.
4. *Hybrid Methods*: Measure the similarity by combining the above ideas.

Information Content (IC) is a measure of specificity for a concept. Higher values are associated with more specific concepts (e.g., pitchfork), while those with lower values are more general (e.g., idea). Information Content is computed based on frequency counts of concepts as found in a corpus of text. The frequency associated with a concept is incremented in WordNet each time that concept is observed, as are the counts of the ancestor concepts in the WordNet hierarchy (for nouns and verbs) [11].

The library WordNet::Similarity[4] implements three Information Content measures: "res" [12], "jcn" [13], and "lin" [14]. The measure proposed in [12] computes the similarity between two concepts as the information content of the most specific concept that both have in common in the is-a hierarchy. The measures proposed in [14] and [13] are both based on the measure proposed in [12]. In our work, we used "lin" measure as it produces a normalized similarity value between 0 and 1 by taking the ratio of the shared information content, explained above in "res" measure, to that of the individual concepts.

Let Sim_Words($W_i$, $W_j$) be the semantic similarity between the words $W_i$ and $W_j$. This measure will be used in our profiles matching algorithm proposed in the next section.

### 5.2 Semantic similarity between profiles

Consider two user's profiles $P_i$ and $P_j$ consisting of n and m words, respectively. To measure the similarity between these profiles, noted Sim_Profiles($P_i$, $P_j$), we proceed as follows:
1. We create n x m matrix corresponding to the words in $P_i$ and $P_j$.
2. We fill in this matrix with the semantic similarity values between each couple of words (cf. section 5.1). Let *Sim_Words($W_{ir}$, $W_{jc}$)* the semantic similarity between the word $W_{ir}$ at row r of profile $P_i$ and the word $W_{jc}$ at column c of profile $P_j$.
3. Let *Sum* and *Counter* be two variables initialized to 0.
4. Find the highest value in the matrix; Let $H_{rc}$ be this value where r and c represents row and column indices. This value represents the best matching between a word from $P_i$ and a word from $P_j$. We mean by best matching in the matrix the most semantically similar words.
5. Let Sum = Sum + $H_{rc}$ ; Let Counter = Counter + 1
6. Assign the value -1 to all the matrix row r and column c to discard them from the coming steps.
7. If all the rows or all the columns of the matrix values are -1, go to step 8, otherwise repeat from step 4.
8. We reach this step because all the matrix values are set to -1. The semantic similarity between the profiles $P_i$ and $P_j$ is the average of summed $H_{rc}$, i.e. *Sim_Profiles($P_i$, $P_j$) = Sum/Counter*.

Consider the following example to illustrate the semantic similarity method between profiles. Let P1 and P2 be two profiles each represented by a set of words. We consider that the words of P1 and P2 are extracted through the hashtag segmentation algorithm (cf. section 4).

P1 = {Information, Office}
P2 = {Salary, Work, Company}

According to the steps 1 and 2 of the proposed algorithm, we construct a matrix filled with the semantic similarity scores between words (*Sim_Words* explained above). We obtain the following matrix:

---

[4] http://wn-similarity.sourceforge.net/

|  |  | **Words of profile P1** | |
|---|---|---|---|
|  |  | Information | Office |
| **Words of profile P2** | Salary | 0.127 | 0.109 |
|  | Work | 0.411 | 0.781 |
|  | Company | 0.388 | 0.615 |

In the above matrix, we can note that the highest semantic similarity is between the words *work* and *office* (value = 0.781).

According to the steps 4 to 7 of the proposed algorithm, we find the indices (r, c) of the cell containing the highest value. We associate the word of P2 on the row r to the word of P1 on the column c. Then, we replace all the cells on the row r and all the cells on the column c by -1. The process is repeated until all the matrix rows and columns become -1. In the above example, the first best matching is $H_{22} = 0.781$ (row 2 and column 2). This means that the word *Work* is best matched with the word *Office*. After setting the values of the row 2 and column 2 to -1, we obtain:

|  |  | **Words of profile P1** | |
|---|---|---|---|
|  |  | Information | Office |
| **Words of profile P2** | Salary | 0.127 | -1 |
|  | Work | -1 | -1 |
|  | Company | 0.388 | -1 |

We find again the best matching from the resulting matrix. Hence, the second best matching is $H_{31} = 0.388$. The word *Company* is best matched with the word *Information*. The matrix becomes:

|  |  | **Words of profile P1** | |
|---|---|---|---|
|  |  | Information | Office |
| **Words of profile P2** | Salary | -1 | -1 |
|  | Work | -1 | -1 |
|  | Company | -1 | -1 |

The resulting matrix has all its values set to -1. We break the algorithm and we go to step 8. The final semantic similarity between the profiles P1 and P2 is then computed as follows:

Sim_Profiles(P1 , P2) = (0.781+0.388)/2 = 0.584

The method proposed in this section will be used in the data clustering explained in the next section.

## 6. Clustering Profiles and Evaluation

By partitioning the profiles into clusters (groups) we can discover related profiles, i.e. users that share common interests. Clustering methods can be applied on a dataset of profiles in order to partition them into clusters such that profiles in the same cluster are more similar to each other than profiles in different clusters according to the matching algorithm explained in the previous section.

We used the Twitter dataset from the Stanford Large Network Dataset Collection[5]. The Twitter dataset in this project consists of 81306 profiles. Each profile consists only of a list of hashtags, i.e. the hashtags cited by the corresponding user. The data is anonymously collected where each profile is represented by an identifier.

We applied our hashtag segmentation algorithm proposed in section 4 on the 81306 profiles. We obtained for each profile a set of significant words. This step took around 20 hours of mono-thread processing on 2.3 GHZ CPU core.

We randomly selected 10000 profiles in order to perform the clustering step. We computed a similarity matrix between the selected profiles. Each value in this matrix represents the similarity between two profiles based on the profile matching algorithm proposed in section 5. We applied the k-medoids clustering method [15]. This is a partitioning technique that clusters the dataset into k clusters. We applied it with 3 different values for k (30, 40 and 50 clusters). The following figures show the profiles distributions in the clusters. For example, in figure 2, the cluster number 26 contains the highest number of profiles (~1200 profiles). The number of clusters is chosen randomly but the number k can be increased or decreased based on some criteria such as to obtain compact clusters.

For a given profile, the system can recommend it some of the most similar profiles in the same cluster. However, the system can construct the social graph for the profiles in the same cluster and recommend profiles as in the traditional methods of the literature but in the same cluster.

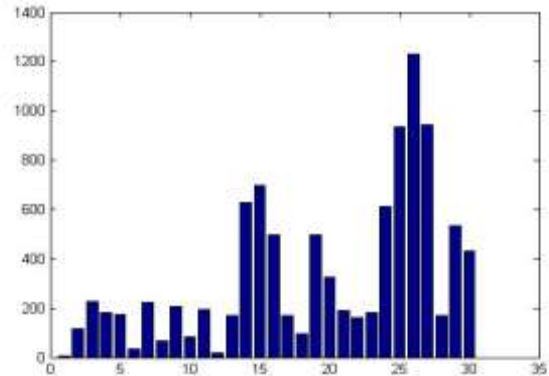
Figure 2: Number of profiles in each cluster obtained by k-medoids with k=30.

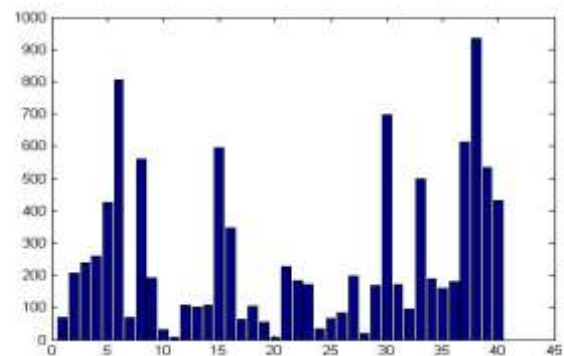
Figure 3: Number of profiles in each cluster obtained by k-medoids with k=40.

---
[5] http://snap.stanford.edu/data/index.html

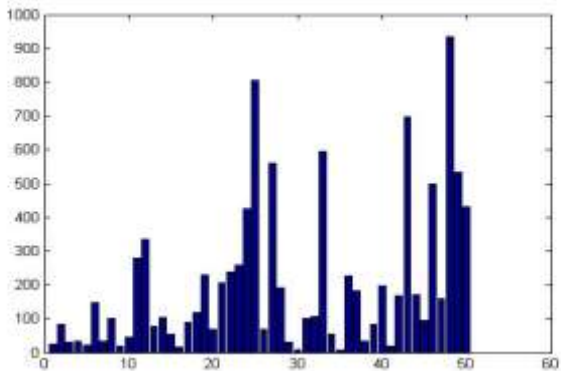

Figure 4: Number of profiles in each cluster obtained by k-medoids with k=50.

## 7. Conclusion

We presented a new approach for recommending friends in social networks. Our approach is based on the semantic content of Hashtags. We implemented a framework that allows constructing user's profiles by extracting hashtag tokens. A semantic matching method is than used to compute the similarity between profiles. Clustering techniques are finally implemented in order to group together semantically close users. Users in the clusters share some interests based on the contents they publish on social networks. Hence, these clusters can be used as the basis of any social recommender system in order to suggest for each user, the top k-neighbors from the cluster she/he belongs to.

As future work, we aim to integrate the hashtag feature to a recommender system that takes into account content attributes defined by the FOAF project. Moreover, we plan to combine hashtag feature and social graphs of profiles in the same framework in order to exploit all valuable information.